\documentclass[10pt,conference]{IEEEtran}
\usepackage{graphicx} % Required for inserting images
\usepackage[utf8]{inputenc}
\usepackage{cite}
\usepackage{algorithm}
\usepackage{amsmath, amssymb}
\usepackage{subfiles}
\usepackage{url}
\usepackage{hyperref}
\usepackage{booktabs}
\usepackage{multirow}
\usepackage{graphicx}
\usepackage{caption}
\usepackage{subcaption}
\usepackage{textcomp}
\usepackage{bm}
\usepackage{xcolor}
\usepackage{amsthm}
\usepackage{algpseudocodex}
\usepackage{arydshln}
\usepackage{pifont}
\usepackage{todonotes}
\usepackage{parskip}

\usepackage[draft, commentmarkup=uwave]{changes}
\definechangesauthor[name={Ankit Kulshrestha}, color=red]{AK}

\newif\ifisjournal
\title{On the importance of hyperparameters in initializing parameterized quantum circuits}
\author{\IEEEauthorblockN{Ankit Kulshrestha}
\IEEEauthorblockA{\textit{Fujitsu Research of America} \\
Santa Clara, CA\\
akulshrestha@fujitsu.com}
\and
\IEEEauthorblockN{Sarvagya Upadhyay}
\IEEEauthorblockA{
\textit{Fujitsu Research of America}\\
Santa Clara, CA \\
supadhyay@fujitsu.com}}

\newcommand{\ket}[1]{|#1 \rangle}
\newcommand{\bra}[1]{\langle #1 |}
\newcommand{\braket}[1]{\langle #1 \rangle}

\newcommand{\Fmn}{\mathcal{F}_{\mu\nu}}
\newcommand{\Mc}[1]{\mathcal{#1}}
\newcommand{\bt}{\bm{\theta}}
\newcommand{\bl}{\bm{\lambda}}
\newcommand{\be}{\bm{\epsilon}}
\newcommand{\bz}{\bm{\zeta}}
\newcommand{\Beta}{\text{Beta}}
\newcommand{\Unif}{\text{Uniform}}

\begin{document}

\maketitle
\begin{abstract}
There has been intensive research on increasing the utility and performance of Parameterized Quantum Circuits (PQCs) in the past couple of years. Owing to this research, there are now several inductive biases available to a quantum algorithms researchers to design a good circuit for their chosen task. 

In this paper, we focus on the problem of finding  performant initial parameters for a given PQC. Different from previous research that focuses on finding the right \emph{distribution}, we focus on finding the \emph{hyperparameters} for any given distribution. To that end we introduce an evolutionary-search based algorithm that finds optimal hyperparameter given a PQC and quantum task. Our empirical results indicate that our algorithm consistently leads to selection of performant initial parameters tuned specifically to the ansatz and the quantum task leading to faster convergence and performance. More importantly, our algorithm does not \emph{negatively} affect the barren plateau phenomenon. In other words, the initial parameters suggested by algorithm do not worsen the gradient variance scaling for a given initializing distribution. 
% Training parameterized quantum circuits (PQCs) has been the subject of intensive research in the past couple of years. Several methods have been proposed to enhance trainability and expressiveness of these types of circuits. One particularly effective method is to focus on finding good initial parameters. 

% Methods in the literature that concern themselves with parameter initialization often focus on finding a good distribution or good constraint on the domain of the distribution so that the circuit does not end up being a one or two design. However, little to no attention is paid to the \emph{hyperparameters} of the  distribution itself. In this paper, we show that choosing hyperparameters of the initializing distribution has a correlation with the trainability of the PQC. We then provide an algorithm to estimate the optimal hyperparameters given a quantum task and the corresponding circuit \emph{without} requiring any additional data.
\end{abstract}

\section{Introduction}
The quest for a workable quantum computer continues to drive both theoretical practitioners and engineers to overcome formidable obstacles in the way. In a few short years, the field has gone from being a theoretical possibility to a practical implementation. While the industry is actively focusing on developing techniques for so-called Early Fault Tolerant Quantum (EFTQC) devices, there has been tremendous development of hybrid quantum-classical algorithms that can run on so-called Noisy Intermediate-Scale Quantum (NISQ) devices. These algorithms called Variational Quantum Algorithms (VQAs) involve running parameterized quantum  circuits (PQCs) whose parameters are optimized on classical computers. 

Designing PQCs is both an art and science. There are many inductive biases that an algorithm designer can choose to create an effective PQC for a given quantum computing task. Some examples of inductive bias include the choice of rotation gate, number of control gates, entanglement patterns, etc. One important inductive design is that of the initializing distribution for the parameters of the circuit. The choice of initializing distribution has a subtle impact on the performance of the circuit. McClean~\emph{et al}~\cite{mcclean2018barren} show that for any Haar random distribution we can expect the circuit to be in a flat landscape even for small sizes. Mitigation strategies for this effect from an initialization perspective have been studied in the literature~\cite{grant2019initialization, kulshrestha2022beinit, zhang2022escaping, park2023hamiltonian}. Through these works it has been gradually established that distributions other than (unmodified) uniform often work better for quantum circuits. One important question that arises from studying initializing distributions is whether we can find good ``hyperparameters" (e.g. $\alpha, \beta$ for Beta, $\mu, \sigma$ for Gaussian) for a given initializing distribution and a given ansatz? Surprisingly, we did not find any literature that deals with this specific question.

% However when choosing a distribution other than uniform, one often runs into a subtle inductive bias - what ``hyperparameters" of the distribution (e.g. $\alpha, \beta$ for Beta, $\mu, \sigma$ for Gaussian) should one choose given a circuit? 

\begin{figure*}[t]
    \centering
    \begin{subfigure}{0.45\textwidth}
        \includegraphics[width=\linewidth]{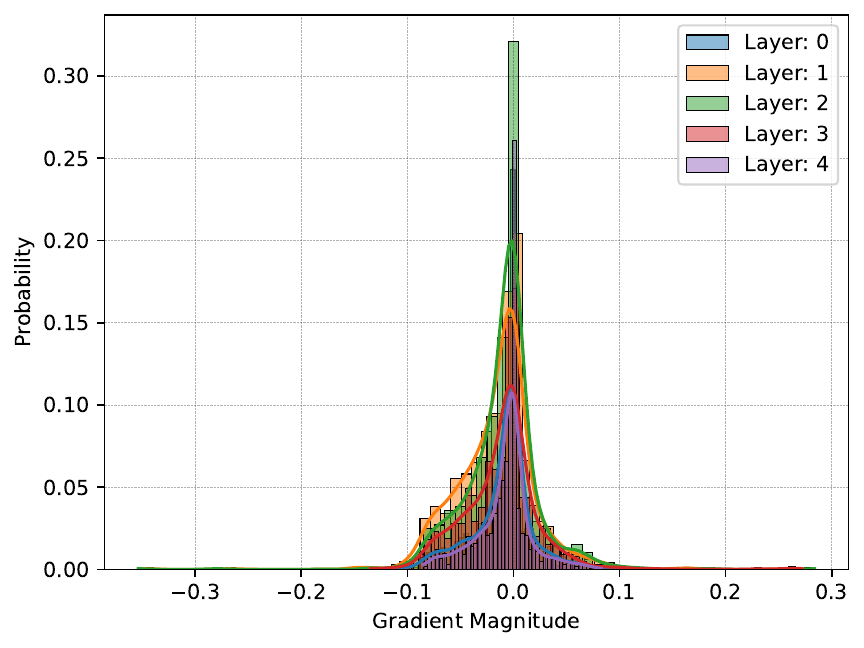}
        \caption{$\bt \sim \text{Beta}(\alpha=0.1, \beta=1.5$)}
    \end{subfigure}
    \hfill
    \begin{subfigure}{0.45\textwidth}
        \includegraphics[width=\linewidth]{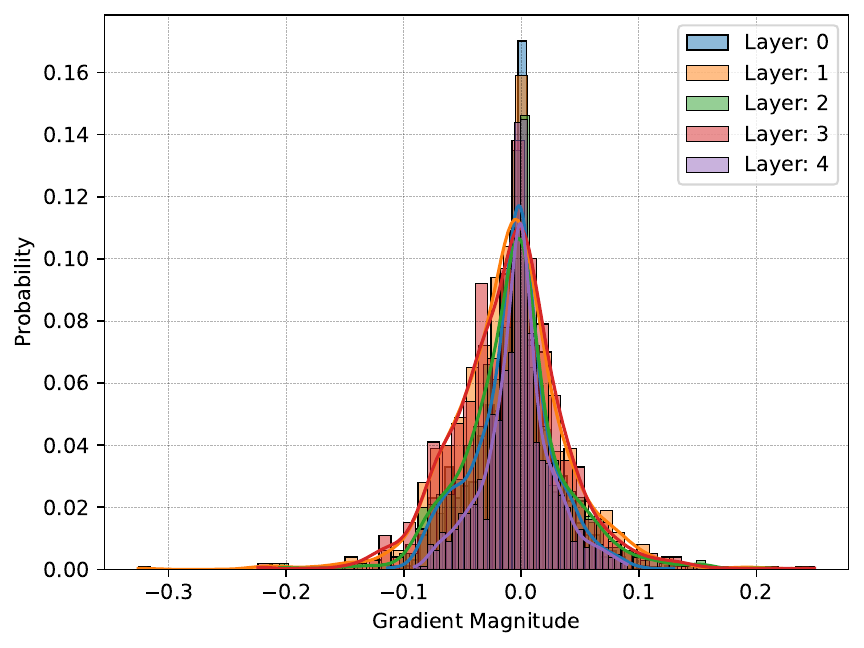}
        \caption{$\bt \sim \text{Beta}(\alpha=0.1+\delta, \beta=1.5+\delta$)}
    \end{subfigure}
    \hfill
    \vspace{1em}
    \begin{subfigure}{0.45\textwidth}
        \includegraphics[width=\linewidth]{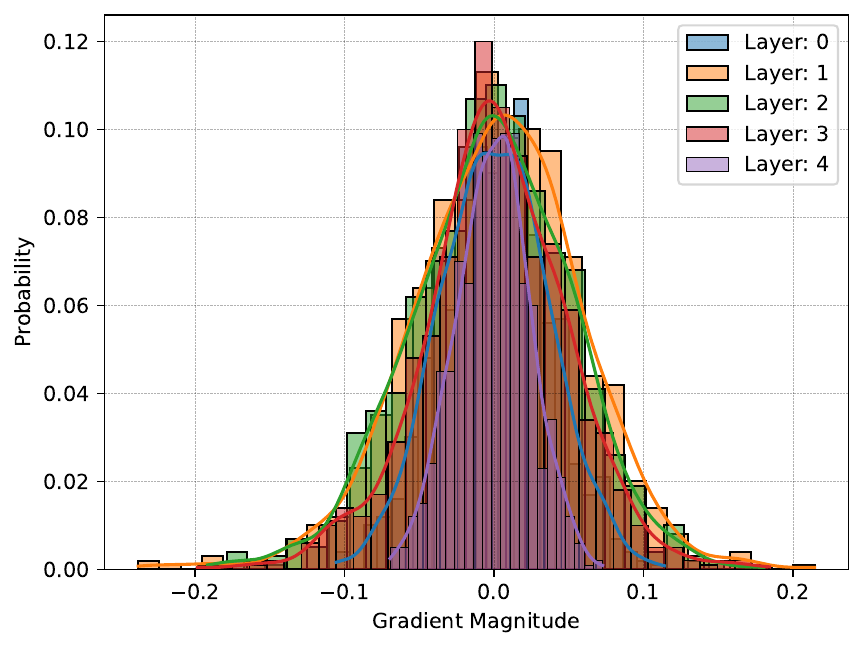}
        \caption{$\bt \sim \mathcal{N}(\mu=0.0, \sigma=0.5$)}
    \end{subfigure}
    \hfill
    \begin{subfigure}{0.45\textwidth}
        \includegraphics[width=\linewidth]{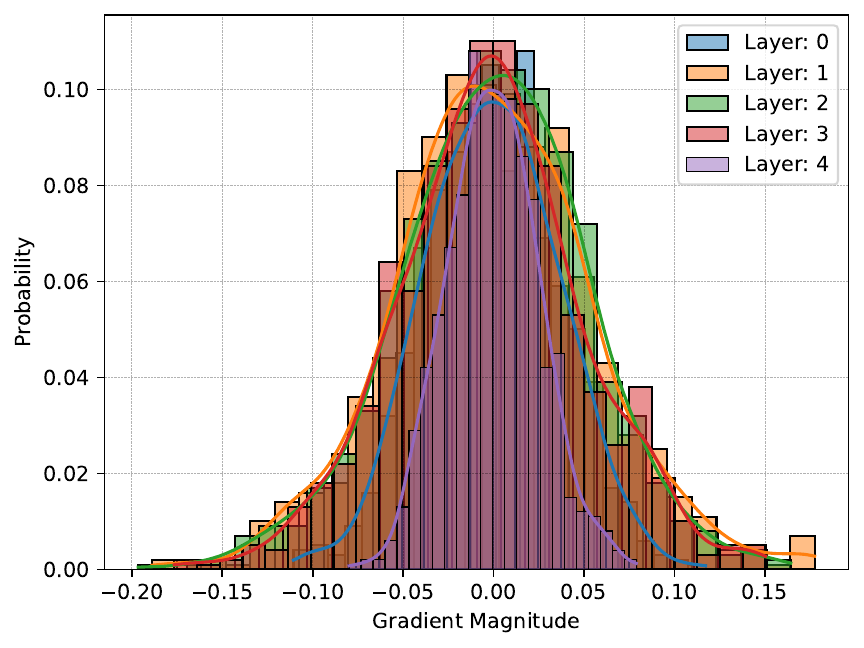}
        \caption{$\bt \sim \mathcal{N}(\mu=0.0 + \delta, \sigma=0.5+\delta$)}
    \end{subfigure}
    \caption{Normalized histogram of gradient magnitude distribution of initial parameters across different layers in a 5 layer, 4 qubit Hardware Efficient Ansatz (HEA). The hyperparameters for the distribution are perturbed by $\delta = 0.05$. The figures show that even a small change in hyperparameters of initializing distribution can lead to drastically different gradient distribution across layers of a quantum circuit.}
    \label{fig:hyp_importance}
\end{figure*}

To establish the importance of hyperparameters, we performed a small experiment with a five layer and four qubit Hardware Efficient Ansatz (HEA)~\cite{kandala2017hardware}. Specifically, we profiled the gradient distribution when parameters are drawn from Gaussian and Beta distributions. The distributions are shown in Figure~\ref{fig:hyp_importance}. The top panel shows gradient distribution with manually selected hyperparameters and the bottom panel shows the distribution when the hyperparameters are perturbed by a small $\delta$. There is a drastic change between the panels in the gradient distributions. The drastic difference is an indication that for parametric distributions, hyperparameter selection is an important and often overlooked inductive bias. 

Motivated by this experiment, we asked whether there was a way to find optimal hyperparameters for any given ansatz and quantum task. To answer this question, we outline an evolutionary search (ES) driven algorithm that finds optimal hyperparameters given an ansatz and a quantum task. We further study the effectiveness of different ``score functions" (also presented in this work) for different distributions and task settings. In brief, we make the following contributions in our paper: 

\begin{itemize}
    \item We study the role of initializing distribution hyperparameters and their subtle contribution to overall performance of PQCs across different quantum tasks.

    \item We propose an evolutionary search based algorithm to estimate hyperparameters for any given PQC and quantum task. Our algorithm is engineered to consume less resources and be fast with access to parallel programming. 

    \item We also study the impact of finding optimal hyperparameters on the barren plateau phenomenon in PQCs. 
\end{itemize}

% \sarvagya{Our contributions: it is important to add contributions here; precise one line statement for each contribution for readers to know what they'll see next}

\section{Hyperparameter Optimization Algorithm}

\subsection{Score Functions}
Hyperparameter search is an indirect optimization problem in the sense that we cannot directly choose the best hyperparameter through a cost function. This leads us to the concept of ``score functions". Abstractly, a score function $\mathcal{S}: \mathbb{R}^d \mapsto \mathbb{R}$ is a class of functions that assign scalar values indicating the utility of the chosen hyperparameters. 

The role of score functions in general is to help us  assess the usefulness of the input. In classical machine learning, the score functions are often modeled as the negative log likelihood loss function. In the case of PQCs however, there are two tightly coupled components. The first is the parameterized circuit which produces an output $\rho({\bm{\theta}})$ and the second is the chosen observable $O$. If we choose the obvious score function, i.e., $\mathcal{S}(\bt) = Tr[\rho(\bt)O]$  then we implicitly condition the score on both the \emph{chosen parameters} and \emph{the choice of the observable}. Clearly, there are infinitely many ways to combine different hyperparameters and observables. Any search algorithm cannot clearly terminate in reasonable time for such a combinatorial search space.

% \ak{Write this passage better}
We propose to utilize a score function that can be tied to the \emph{ansatz} instead of the choice of observable. This reduces the complexity of search space and enables a search algorithm to finish in reasonable time. We can additionally pair such a score function with some metric that quantifies the performance on the given quantum task. The combined information then represents an interpolation between the inductive biases specific to the ansatz and it's performance given a particular observable and cost function. One attractive metric for the ansatz specific score function is the Quantum Fisher Information Matrix(QFIM)~\cite{petz2011introduction}. For a pure parameterized state $\ket{\psi(\bt)}$, the QFIM is four times the Quantum Metric Tensor(QMT):
% Motivated by this observation, we propose a modular score function that can convey information about the structure of the circuit and the performance on the given quantum task. Help for the first component comes in the form quantum Fisher Information matrix (QFIM)~\cite{petz2011introduction}\sarvagya{Citation missing and this sentence seems a bit awkward}. For pure quantum states $\ket{\psi}$, the QFIM is four times the quantum metric tensor:\sarvagya{The bra-ket thing with $\partial$ makes it confusing to understand the equation.} 

\begin{equation}\label{eq:qfim}
    \Fmn(\bt) = 4Re\big(\braket{\partial_\mu \psi | \partial_\nu \psi} - \braket{\partial_\mu \psi | \psi}\braket{\psi | \partial_\nu \psi}\big)
\end{equation}

Where $\ket{\partial_\mu \psi} = \frac{\partial}{\partial \theta_\mu}\ket{\psi(\bt)}$ is the derivative w.r.t the $\mu^{th}$ component of the parameter vector and $\ket{\psi}$ represents the parameterized state $\ket{\psi(\bt)}$. For a mixed state $\rho(\bt)$, the QFIM is:

% In the general case, the QFIM is derived as \sarvagya{Are we working with the general case? If not, we can skip it.}
\begin{align}\label{eq:qfim_density}
    \mathcal{F}_{\mu\nu}(\bt) & = \frac{1}{2}Tr[\rho(\bm{\theta})\{\mathcal{L}_\mu, \mathcal{L}_\nu\}] \nonumber \\ 
    & = \sum_{m,n (\lambda_m + \lambda_n > 0)} 2 \frac{Re\big(\bra{m}\partial_\mu \rho \ket{n} \bra{m}\partial_\nu \rho \ket{n}\big)}{\lambda_m + \lambda_n}
\end{align}
Here $\lambda_m, \lambda_n$ are the eigenvalues in the spectral basis of $\rho(\bm{\theta})$. The key property of QFIM (that we are interested in) is that for unbiased estimators $\hat{\bm{\theta}}$:
\begin{equation}\label{eq:cr_bound}
    Cov(\hat{\bm{\theta}}) \geq (\mathcal{F}_{\mu\nu}^{O_x})^{-1} \geq (\mathcal{F}_{\mu\nu})^{-1}
\end{equation}
 
Where $\Fmn^{O_x}$ is the QFIM associated with an observable $O_x \in \{O_1, O_2 \dots O_N\}$. The relation tells us that $\Fmn = \sup_{O} \Fmn^{O}$. The upshot of the discussion is this: \emph{QFIM allows us to estimate the sensitivity of the circuit to the given parameters irrespective of the choice of observable}. 

However, QFIM itself may not always provide enough information by itself, since for $\bm{\theta}' = \bm{\theta} + \delta$ and $\delta <<< 1$, $\Fmn(\bm{\theta}') \approx \Fmn(\bm{\theta})$. Thus,for a given task we may need to pair this score with a \emph{task specific score}. In this work, we propose three different score functions:
% \todo[inline]{Experiment with Harmonic Trace Score function}
\begin{align}\label{eq:score_fn}
    \Mc{S}_1(\bt) &:= \Omega(\Fmn(\bt))\\
    \Mc{S}_2(\bt) &:=  \Mc{M}^t(\nabla C(\bt))\\
    \Mc{S}_3(\bt) &:= (1 - w) .\Omega(\Fmn(\bt)) + w . \Mc{M}^t(\nabla C(\bt))  
\end{align}

Here, $\Mc{M}^t(\nabla C(\bt))$ is the $t^{th}$ order statistic on the gradient of the cost function defined for the task. In most of our experiments we set $t=2$ and $\Omega(.)$ is a scalar reduction function for the input QFIM. Typically we set $\Omega(.)$ as the trace operator. We can also use the Log-Determinant as well. The three different score functions serve distinct purposes in this study. We have already discussed the motivation for the first one in great detail earlier. The second score function studies the contrastive effect of just focusing on order statistics of the given quantum task (e.g. VQE, QAOA, etc.). The final one is a convex combination between these two extremes and aims to study if a middle ground exists that can lead to better hyperparameter suggestions.

\subsubsection{Implementation Details}
The complexity of computing QFIM exponentially increases with an increase in the number of parameters. To implement a practical algorithm, we compute a ``block-diagonal" approximation to the QFIM and for large parameter sizes default back to computing the so-called ``empirical QFIM" i.e. $\nabla C(\bt)\otimes \nabla C(\bt) \in \mathbb{R}^{p \times p}$ where $p = |\bt|$. 

Additionally, score function $\Mc{S}_1$ in Equation~\ref{eq:score_fn} requires careful implementation to prevent exploding/collapsing values. To achieve stable computation we compute $\Mc{S}_1(\bt) = \Omega(\Fmn(\bt) + \epsilon. \bm{I})$. 

\subsection{Estimation Algorithm}

The hyperparameters of a given parameter distribution can be estimated by several off-the-shelf algorithms that exist in literature. For instance, one can use a Bayesian Optimization (BO)~\cite{bergstra2011algorithms} approach to estimate the GP posterior for the given set of hyperparameters. One can also define MLE estimators for the distribution in closed form and use it to estimate the optimal hyperparameters. There are also simpler options like Grid search and Random Search. 

However, these algorithms are proposed in the context of classical machine learning. The case of PQCs is both subtle and challenging owing to the owing to the overhead of the calls that must be made to the QPU. Table~\ref{tab:hyp_complexity} summarizes the algorithmic runtime complexities of the aforementioned algorithms. From the table, it is clear that many sophisticated algorithms incur a high quantum resource usage. Meanwhile, simpler algorithms like Random Search save on expensive inner loops but become expensive owing to the number of iterations that need to be run for getting a good estimate. 

% \todo[inline]{add stuff describing the table}
\begin{table}[h]
    \centering
    \begin{tabular}{c|c | c}
         Method & Algorithmic Complexity & Parallelizable?\\
         \hline
         Bayesian Optimization  &  $O(MN^2C_q) $ &\ding{55}\\ 
         MLE Estimation & $O(TN_{samp}C_q)$ & \ding{55}\\ % check accuracy of this one
         Random Search &  $O(TC_q)$&  Partial \\ 
         Grid Search & $O(n^dC_q)$ & \ding{55}\\ 
    \end{tabular}
    \caption{Algorithmic complexities of existing hyperparameter search algorithms. In BO, $M$ is the number of acquisition function steps, $N$ is GP evaluations. For MLE, $T$ is the total number of iterations and $N_{samp}$ is the number of samples required per iteration to get a low variance estimate. For Grid search, $n$ is the number of hyperparameters and $d$ is their dimensionality. Assuming gradients/QFIM is computed on a quantum device using a statevector model $C_q \approx O(p^22^q)$ for $\bt \in \mathbb{R}^p$ parameters.}
    \label{tab:hyp_complexity}
\end{table}
% It is inspired by related work in evolutionary search strategies~\cite{Salimans, Weistra}

The shortcomings of the classical algorithms when applied to the quantum case motivated us to search for other methods. Evolutionary Search (ES)~\cite{salimans2017evolution, wierstra2014natural, sun2009efficient} algorithms are frequently used in reinforcement learning scenarios especially in cases where gradient computation and backpropagation is expensive. Evolutionary search based methods have the advantage of modeling complex black box optimization functions. Moreover, they do not require any posterior update like BO or iterative minimization like MLE. The most attractive feature of this method type is that it is \emph{embarrassingly parallel}. Thus even though the serial version has a complexity of $O(T\Lambda C_q)$ where $\Lambda$ is the population size, the parallel version has a complexity of $O(T\frac{\Lambda}{N_{cores}} C_q)$ where $N_{cores}$ is the number of cores available on the device. For large enough cores, $\frac{\Lambda}{N_{cores}} < 1$.

\begin{algorithm}
\caption{ES-HyperOpt Algorithm}
\label{alg:hyp_opt}
\begin{algorithmic}[1]
    \Require $\Mc{S}$: Score function, $\bl_0$: initial distribution hyperparameter guess, $p(\bt | \bl)$: Initializing distirbution, $\eta$: Learning rate, $\sigma$: step size, $N_s$: Num samples per iteration, $N_{iters}$: Number of iterations
    \Ensure Optimal hyperparameters $\bl^*$
    \State $i \gets 0$
    \State $\bz \gets [\varnothing]$
    \While{$i < N_{iters}$}
       \State $\bm{\Gamma} \gets \Mc{N}(0, 1) \in \mathbb{R}^{N_s \times |\bl|}$
       \For{$j = 0, 1, \dots, N_s$}
            \State $\bl^p \gets \bl + \be_j$ \Comment{$\be_j \in \Gamma$ is $j^{th}$ perturbation vector}
            \State $\bt^p \sim p(\bt | \bl^p)$
            \State $\text{score}_j \gets \Mc{S}(\bt^p)$
            \State $\bz \gets \bz \cup \text{score}_j$
       \EndFor
       \State $\nabla_{\bm{\lambda}} = \frac{1}{N_s.\sigma}(\bz^T \Gamma)$
       \State $\bl' \gets \bl + \eta \nabla_{\bm{\lambda}}$ \Comment{Gradient ascent step}
       \State $i \gets i +1$
    \EndWhile
    \Return $\bm{\lambda}^{*}$ 
\end{algorithmic}
    
\end{algorithm}

Algorithm~\ref{alg:hyp_opt} shows our proposed approach. We input a chosen score function and an initial guess ($\bl_0$) for a  parameterized distribution (e.g. Gaussian, Beta etc). There are two distinct stages in the algorithm. The first stage is the ``rollout" stage where we perturb hyperparameters by a noise perturbation drawn from $\Mc{N}(0,1)$ (Lines 5-7). From these perturbed distributions, we draw parameter samples $\bt \sim p(\bt | \bl')$. Then, the score function is used to estimate the scalar score for the given quantum circuit and the task. These returns are then handed to the second stage (Lines 8-9) which computes the gradient and the update to the hyperparameters. The algorithm terminates if $||\bl_{i+1} - \bl_{i}||_1 \leq \varepsilon_{converge}$. Since the perturbations can be independently sampled they tend to have a high variance. This can often hinder convergence, especially if the dimensionality of the hyperparameters is high. To control variance, we employ antithetic sampling (also used in related works~\cite{salimans2017evolution, wierstra2014natural}). In this scheme we sample $\be$ for $\frac{N_s}{2}$ size. For the remaining half, we employ $-\be$ perturbations. Since $\be, -\be$ are negatively correlated, the joint variance between these two random variables is overall decreased. 
\begin{figure*}[t]
    \centering
    \begin{subfigure}{0.45\textwidth}
        \includegraphics[width=\linewidth]{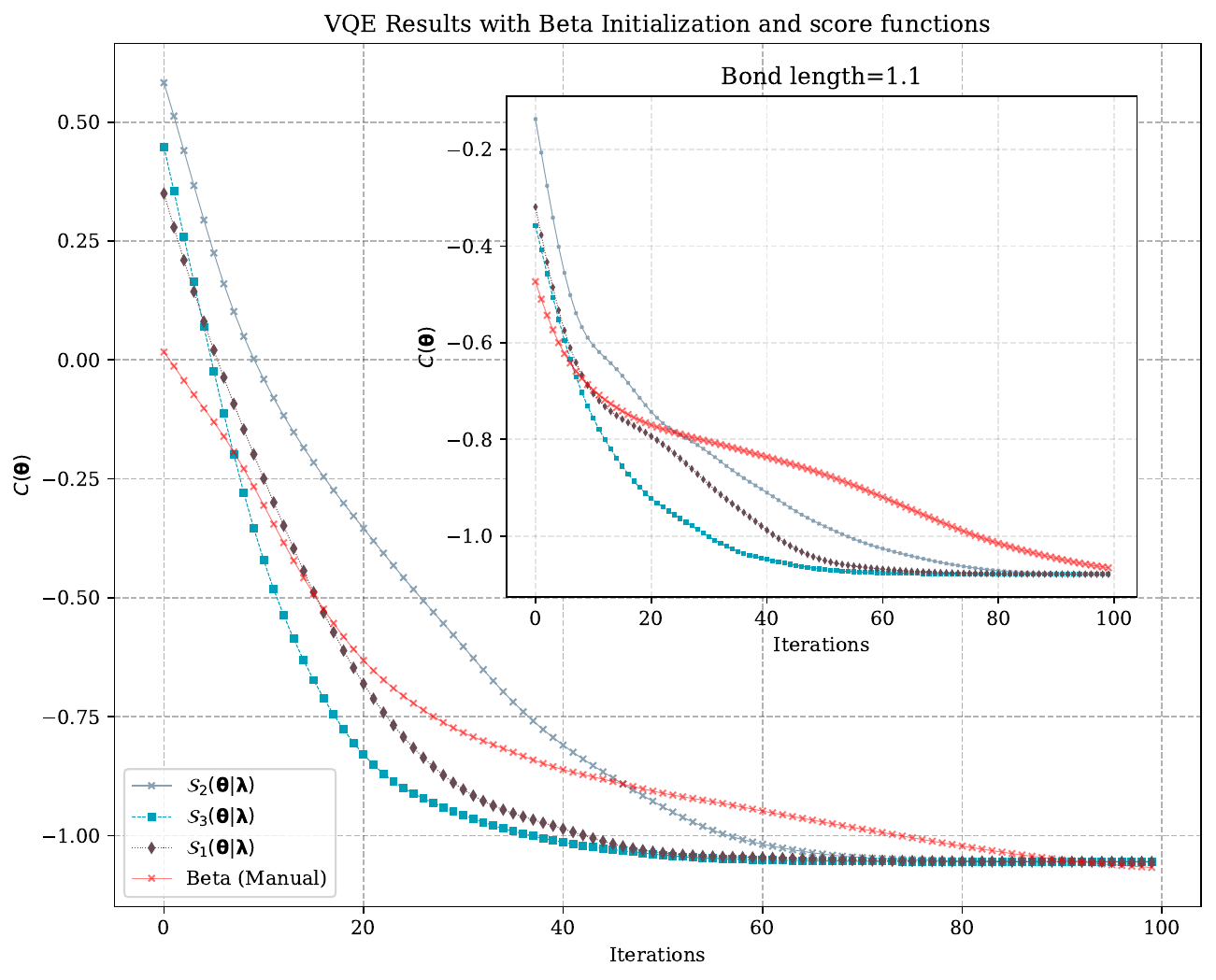}
        \caption{Beta Distribution}
        \label{sfig:beta}
    \end{subfigure}
    \hfill
    \begin{subfigure}{0.45\textwidth}
        \includegraphics[width=\linewidth]{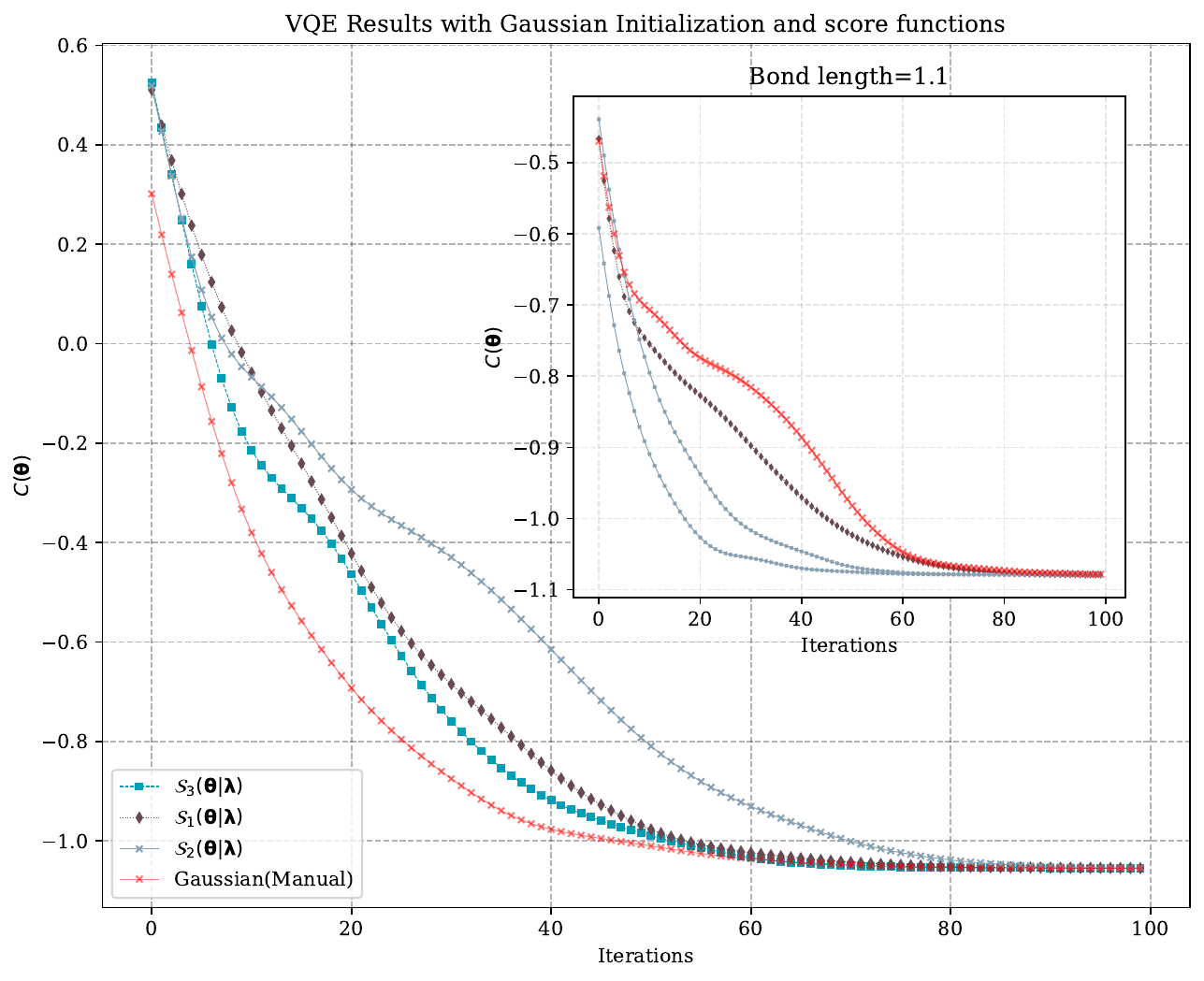}
        \caption{Gaussian Distribution}
        \label{sfig:gaussian}
    \end{subfigure}
    \hfill
    % \vspace{1em}
    % \begin{subfigure}{0.45\textwidth}
    %     \includegraphics[width=\linewidth]{Figures/grad_stats_gauss_0.0_0_5.pdf}
    %     \caption{$\bt \sim \mathcal{N}(\mu=0.0, \sigma=0.5$)}
    % \end{subfigure}
    % \hfill
    % \begin{subfigure}{0.45\textwidth}
    %     \includegraphics[width=\linewidth]{Figures/grad_stats_gauss_0_05_0_55.pdf}
    %     \caption{$\bt \sim \mathcal{N}(\mu=0.0 + \delta, \sigma=0.5+\delta$)}
    % \end{subfigure}
    \caption{VQE Training results for $H_2$ molecule with bondlength $\in [0.5, 1.1]\mathring{A}$ for Beta and Gaussian Distributions. The results show that searched hyperparameters with the given score functions produce a faster convergence in general.}
    \label{fig:vqe_results}
\end{figure*}
Given a circuit that is robust to small parameter perturbations, the score function values $\Mc{S}_1, \Mc{S}_2$ may produce raw scores that don't provide a strong gradient signal. Instead of using raw scores, we can construct a rank-based scoring function. Consider a population $\Lambda$ with individual scores ordered as $\{z_1 \geq z_2 \geq \dots z_{\Lambda}\}$. This ordering serves as a relative ranking between different members of the population. We can then derive a \emph{utility} score that assigns each member of the population a utility based on their rank. For the individual with the $k^{th}$ rank in the population we derive the utility score as:
\begin{equation}\label{eq:fitness_shaping_openai}
u_k = \frac{k}{(N_s-1)} - 0.5
\end{equation} 

This utility function is used in~\cite{salimans2017evolution} and has a robust theoretical motivation in related works~\cite{wierstra2014natural, sun2009efficient}. In our experiments, utility based score derivation led to faster convergence of the search algorithm. 

% \ak{This passage ends abruptly, maybe have a neater segue into experiments.}
% % The use of relative rank and utility functions is well known in literature~\cite{wierstra2014natural, sun2009efficient, salimans2017evolution}. One big advantage of using this meth
% % The utility is thus clamped between $[-0.5, 0.5]$. The advantage of this utility score is that it uses relative rank instead of raw scores and thus will produce large enough gradient estimates to allow convergence. This type of score function derivation is termed fitness shaping in the literature~\cite{wierstra2014natural, sun2009efficient, salimans2017evolution}. 

\section{Experiments}
% TODO: add thematic stuff here.

\subsection{Task based evaluation of Hyperparameter Selection Algorithm}
Measuring effectiveness of hyperparameter search algorithm is often an indirect process i.e. we can only regard hyperparameters as ``good" if they lead to a measurable performance gain in the downstream task. To evaluate if our algorithm proposes such ``good" hyperparameters, we measure the performance on two different quantum tasks - estimating the ground state energy of a given molecule (VQE) and classifying data into classes (QML).\\ 

Our experimental setup for both sets of experiments is as follows. We create an ansatz by utilizing \texttt{StronglyEntanglingLayers} structure from PennyLane~\cite{bergholm2018pennylane}. For VQE, the Hamiltonian of the $H_2$ molecule determines the number of qubits and for QML tasks, we perform a PCA projection of high dimensional data into $4$ components. The ansatz itself consists of $8$ such layers. The overall parameter shape for the experiments is thus $(8, q, 3)$ where $q$ denotes the number of qubits in the circuit. In both tasks, the circuits are trained for $100$ iterations with the Adam\cite{kingma2014adam} optimizer with a learning rate $\eta=0.01$. In both sets of experiments we provide results with $p(\bt) \sim \Beta(\alpha, \beta)$ and $p(\bt) \sim \Mc{N}(\mu, \sigma)$. For the Gaussian distribution, we initialize $\mu \sim \Unif(0.1, 0.5)$ and $\sigma \sim \Unif(0.5, 1.0)$. Initializing $\alpha, \beta$ on the other hand is tricky because there is no good upper bound on them. We initialize $\alpha, \beta \sim \log(\Unif(1.0, 5.0))$ and then perform search in the log-space. We find that this prevents extremely large and extremely small values of the optimized hyperparameters. 

For these sets of experiments, we set $\Omega(.) = Tr[.]$ in score functions $\Mc{S}_1, \Mc{S}_3$. Additionally, we use $w=0.9$ for $\Mc{S}_3$.

\subsubsection{VQE Results}

Figure~\ref{fig:vqe_results} shows the results of our VQE experiments. We perform optimization on the $H_2$ molecule with bond length $b \in \{0.5, 1.1\}$. The circuit is trained with parameters drawn from $p(\bt | \bl^*)$ for the three score functions. The manual hyperparameters are set to $\alpha=0.1, \beta=1.5$ following suggestions from~\cite{kulshrestha2022beinit}. Similarly, we set $\mu=0, \sigma=1$ for Gaussian distribution in the manual case. 

From the figure, we can see that the ansatz converges to the $E_{FCI}$ of $H_2$ in both cases. However, the key detail from the figure is the \emph{speed} of convergence. In the case of $H_2^{b=0.5}$ we can see that hyperparameters suggested by both $\Mc{S}_1, \Mc{S}_3$ outperform Manual hyperparameters. This performance is improved in $H_2^{b=1.1}$ (inset Fig~\ref{sfig:beta}) when hyperparameters suggested by all three score functions outperform manually suggested ones. In the case of Gaussian, for small bond length, the manually chosen hyperparameters appear to have a better convergence performance but this significantly changes when larger bond lengths are considered (inset figure) as the manual selection performs considerably worse than all score functions. The results are a strong indication that our algorithm suggests good circuit-specific hyperparameters for the initializing distributions. These hyperparameters in turn help the circuit train faster and attain ground state energy quicker than the baseline method.

% \begin{figure*}[htbp]
%     \centering
%     % \begin{subfigure}{0.45\textwidth}
%     %     \includegraphics[width=\linewidth]{Figures/VQE_Results_Beta_v2.pdf}
%     %     \caption{Beta Distribution}
%     % \end{subfigure}
%     \hfill
%     \begin{subfigure}[b]{0.3\textwidth}
%         \includegraphics[width=1.5\linewidth]{Figures/wine_beta_results.pdf}
%         \caption{Wine Dataset}
%     \end{subfigure}
%     \hfill
%     \begin{subfigure}[b]{0.3\textwidth}
%         \includegraphics[width=1.5\linewidth]{Figures/breast_cancer_beta_results.pdf}
%         \caption{Breast Cancer Dataset}
%     \end{subfigure}

%     \begin{subfigure}[b]{0.3\textwidth}
%         \includegraphics[width=\linewidth]{Figures/digits_beta_results.pdf}
%         \caption{Digits Dataset}
%     \end{subfigure}

%     % \vspace{1em}
%     % \begin{subfigure}{0.45\textwidth}
%     %     \includegraphics[width=\linewidth]{Figures/grad_stats_gauss_0.0_0_5.pdf}
%     %     \caption{$\bt \sim \mathcal{N}(\mu=0.0, \sigma=0.5$)}
%     % \end{subfigure}
%     % \hfill
%     % \begin{subfigure}{0.45\textwidth}
%     %     \includegraphics[width=\linewidth]{Figures/grad_stats_gauss_0_05_0_55.pdf}
%     %     \caption{$\bt \sim \mathcal{N}(\mu=0.0 + \delta, \sigma=0.5+\delta$)}
%     % \end{subfigure}
%     \caption{Training results for a PQC classifier on three different datasets with the Beta Distribution. In all cases, the score function based hyperparameters lead to faster convergence than manually selected hyperparameters.}
%     \label{fig:qml_results}
% \end{figure*}

\begin{figure*}[htbp]
    \centering
    \includegraphics[scale=0.5]{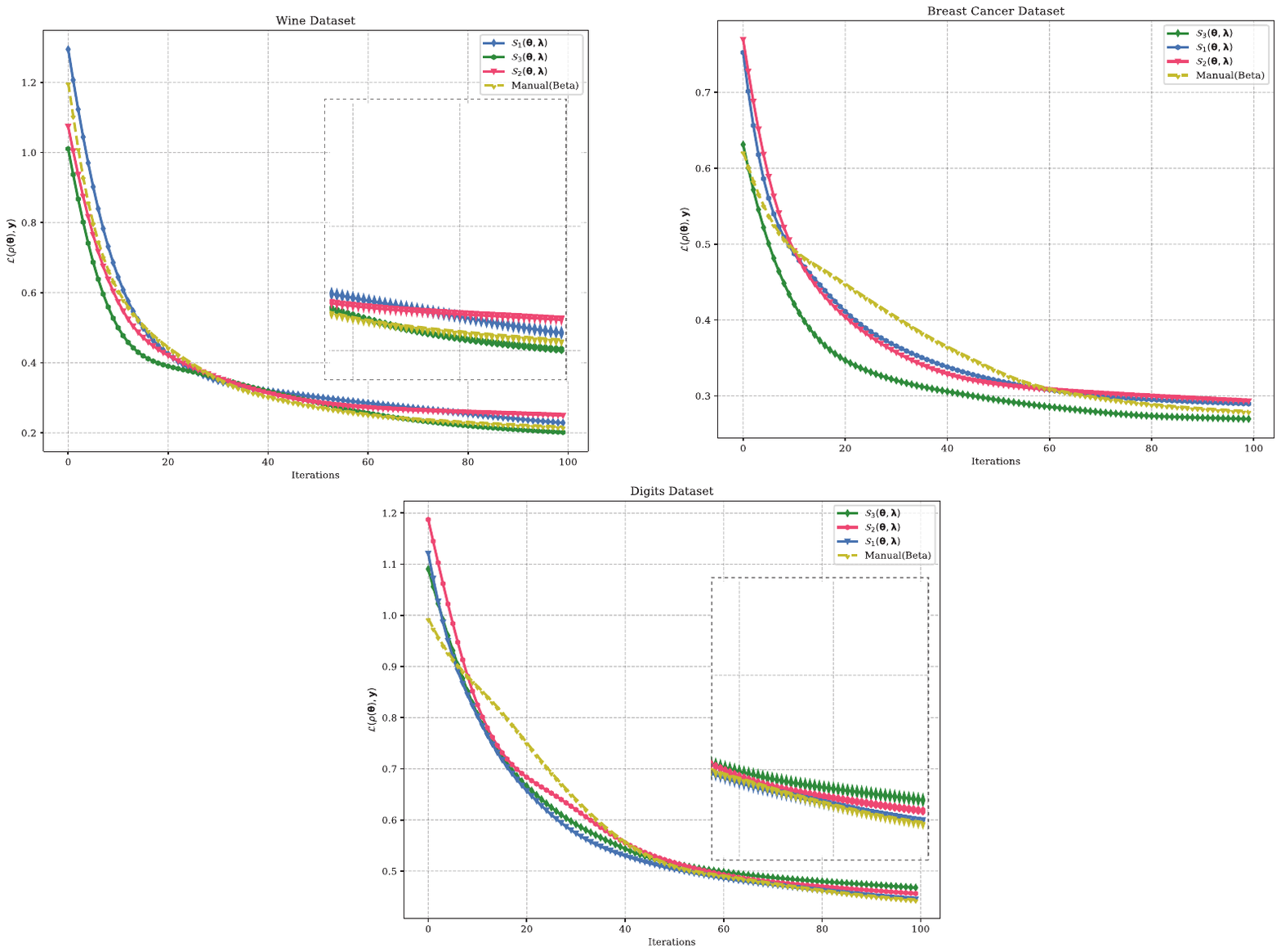}
    \caption{Training loss on different QML datasets with different score functions and manually selected hyperparameters for initializing distributions. Inset: convergence dynamics at the end of training. In all the cases, method based on score function leads to faster convergence than manual selection of hyperparameters even if the ansatz converges to the same (local) minima.}
    \label{fig:qml_results_beta}
\end{figure*}

\subsubsection{QML Results}
\begin{table}[h]
    \centering
    \begin{tabular}{cccc}
        \toprule
        Dataset & $N$  & $d_{in}$ & $d_{pca}$  \\ 
        \midrule
        Wine & $178$ & $13$ & $4$ \\ 
        Breast Cancer & $569$ & $30$ & $4$\\
        Digits & $1797$ & $64$ & $4$\\ 
        \bottomrule
    \end{tabular}
    \caption{The datasets considered in our QML study along with their total dataset size ($N$), input feature dimension $d_{in}$ and PCA dimension $d_{pca}$.}
    \label{tab:data_stats}
\end{table}

\begin{table*}[t]
\centering
\caption{QML Test Accuracy (out of maximum $1.0$) on Different Datasets with Different Initializations and Score Function Choices. \textit{Manual} denotes manually selected hyperparameters per dataset. Bold red indicates best accuracy for beta and bold blue indicates best accuracy for Gaussian Distributions.}
\label{tab:qml_results}
\begin{tabular}{l ccc@{\hspace{0.8em}}c ccc@{\hspace{0.8em}}c}
\toprule
& \multicolumn{4}{c}{$p(\bt | \bl) \sim \Beta(\alpha, \beta)$} 
& \multicolumn{4}{c}{$p(\bt | \bl) \sim \Mc{N}(\mu, \sigma)$} \\
\cmidrule(lr){2-5} \cmidrule(lr){6-9}
\textbf{Dataset} &  $\Mc{S}_1$ &  $\Mc{S}_2$ &  $\Mc{S}_3$ & \textit{Manual} 
                 & $\Mc{S}_1$ &  $\Mc{S}_2$ &  $\Mc{S}_3$ & \textit{Manual} \\
\midrule
Wine &  $\color{red}\bm{0.96}$ & $0.96$ & $0.92$ & $0.92$ & $0.88$ & $0.96$ & $\color{blue}\bm{1.00}$ & $0.80$ \\
Breast Cancer & 0.44 & 0.56 & $\color{red}\bm{0.60}$ & 0.51 & 0.52 & 0.57 & $\color{blue}\bm{0.62}$ & 0.59 \\
Digits & $\color{red}\bm{0.87}$ & 0.86 & 0.83 & 0.79 & $\color{blue}\bm{0.83}$ & 0.79 & 0.79 & 0.79 \\
\bottomrule
\end{tabular}
\end{table*}

%%%% Digits Beta Results %%%% 
% \begin{tabular}{lrr}
% \toprule
%  & train_acc & test_acc \\
% \midrule
% score_beta_qfi & 0.879725 & 0.876712 \\
% score_grad_beta & 0.865979 & 0.863014 \\
% score_grad_beta_hybrid & 0.893471 & 0.835616 \\
% manual & 0.896907 & 0.794521 \\
% \bottomrule
% \end{tabular}

%%% Digits Gaussian Results %%% 
% \begin{tabular}{lrr}
% \toprule
%  & train_acc & test_acc \\
% \midrule
% score_grad_gaussian_hybrid & 0.896907 & 0.794521 \\
% score_grad_gaussian & 0.893471 & 0.794521 \\
% score_gaussian_qfi & 0.879725 & 0.835616 \\
% manual & 0.886598 & 0.794521 \\
% \bottomrule
% \end{tabular}
To further establish the effectiveness of our proposed algorithm, we test it on a classification task using a PQC with three different datasets shown in Table~\ref{tab:data_stats}. We use the same ansatz as in the VQE case study but only measure the output ``logits" by measuring the first qubit in the $\sigma_z$ basis. All datasets are partitioned using an 80/20 train-test split. The PQC is trained independently on each dataset with parameters drawn from a Beta and Gaussian distribution using an Adam~\cite{kingma2014adam} optimizer for $100$ iterations and a learning rate of $0.01$. For the manual case of hyperparameter selection, we utilize the same set of hyperparameters as in the VQE case study. For each dataset, we use angle embedding to embed the data points in quantum space i.e. $\ket{\psi_{in}}_{j} = e^{-if_jW_j}$.

Our evaluation on the QML task takes a multi-metric view. Besides the conventional method of evaluating the trained PQC on the test set, we also evaluate the rate of convergence when compared to the manual selection. Table~\ref{tab:qml_results} shows the test set accuracy of the PQC on the different datasets considered in this study. It is clear that for both Beta and Gaussian distributions our algorithm finds hyperparameters that lead to a dramatic increase in performance when compared to the manual case. Across all datasets we see an average performance increase of $9.3\%$ in the case when parameters are drawn from a Beta Distribution and $12.6\%$ when the parameters are drawn from Gaussian distribution. Empirical evidence further suggests that for the Beta distribution a QFIM-based score function provides enough information to draw good hyperparameters while for Gaussian a hybrid strategy is more effective. Given that the hyperparameters of the Beta distribution are unbounded in $(0, \infty)$, it is plausible that the gradient-based score function is less informative than the Gaussian case where the hyperparameters are generally bounded in $(0, 1]$. 

To evaluate the rate of convergence, we plot the training dynamics of the ansatz under test on all datasets with the same setup as VQE experiments. Figure~\ref{fig:qml_results_beta} shows the results with the Beta distribution. From the figure it becomes clear that in the case when hyperparameters are searched with our proposed score functions, the rate of convergence is much faster than the case when hyperparameters are chosen manually. Note that in cases when the ansatz converges to the same (local) minima, the test accuracy can often be the same for the final parameters (e.g. for the Wine dataset). The plots and tables for this task also help us connect the performance of our method with the size of the dataset. For small datasets (e.g. Wine) the difference in score-based and manual selection is minimal except for the convergence dynamics. However, for medium (e.g. Breast Cancer) to large (e.g. Digits) datasets the score based method significantly outperforms the manual selection method both in terms of convergence dynamics and test set performance.

\subsection{Connection to Barren Plateaus}
\begin{figure}[h]
    \centering
    \includegraphics[scale=0.5]{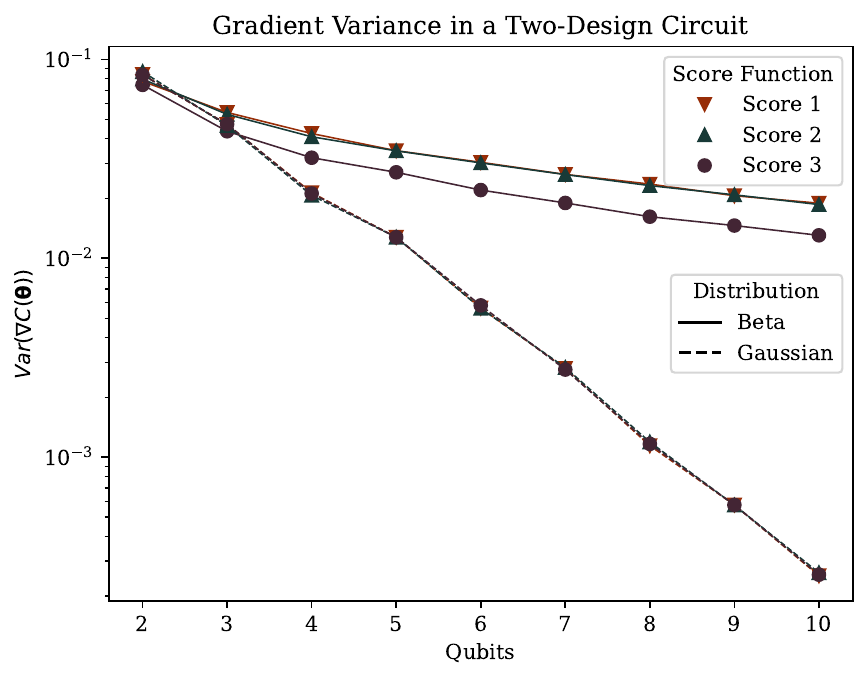}
    \caption{Gradient variance scaling for a two-design exhibiting ansatz for Beta and Gaussian distributions with different score functions.}
    \label{fig:grad_scaling_result}
\end{figure}

We performed a set of experiments to establish the occurrence of barren plateaus in a given ansatz and if the score functions proposed in Equation~\ref{eq:score_fn} have an effect on gradient variance scaling. The experiments were performed with an ansatz that is known to exhibit a 2-design~\cite{mcclean2018barren} and the parameters are initialized from a Beta and Gaussian distributions. For score functions involving the QFIM ($\Mc{S}_1, \Mc{S}_3$), we profiled three different types of $\Omega$. First, we set $\Omega(\Fmn) = Tr(\Fmn)$  as proposed earlier. We also experimented with $\Omega(\Fmn) = \log(|\Fmn + \epsilon I|)$ where $|.|$ denotes the determinant and $\Omega(\Fmn) = K\sum_{i=1}^{k} \frac{1}{\lambda_k}$ where $\lambda_1, \dots, \lambda_k$ denote the $k$ most significant eigenvalues of $\Fmn$. We refer to the first case as ``trace", second as ``log-det" and third as ``harmonic".  We profile the gradient variance for all three score functions $\Mc{S}_1,\Mc{S}_2,$ and $ \Mc{S}_3$ with different types of $\Omega$. Figure~\ref{fig:grad_scaling_result} shows the result when system sizes are varied from $2$ to $10$ for a $5$ layer ansatz. The curves for $\Mc{S}_1$ and  $\Mc{S}_3$ show the best performing $\Omega$ from amongst the choices. 
 
In the figure, Score 1 represents the scaling curve for $\Mc{S}_1$, Score 2 for $\Mc{S}_2$ and so on. For both distributions, we see that the overall scaling is consistent with previously obtained results~\cite{kulshrestha2022beinit}. For the beta case, $\Mc{S}_1$ works best with the ``log-det" choice of $\Omega$ while for $\Mc{S}_3$ the ``harmonic" choice works the best. One reason for this difference is that the space of $\alpha, \beta$ parameters is not upper bounded. Thus, in the hybrid case tracking the eigenvalues can point to the most significant directions in gradient thereby reinforcing the signal obtained from profiling the gradient variances. However, in cases where the range of shape parameters is constrained (e.g. in Gaussian distribution) we see that the effect of score functions is minimal.

The apparent negative result highlights an important point in the paper i.e. we can obtain better performance from our circuit for a given task if we initialize from a distribution whose shape parameters are specially tuned \emph{without} running a risk of leading the ansatz into a barren plateau.

\section{Related Work}

\textbf{Hyperparameter Estimation}: Hyperparameter estimation has a rich history in classical machine learning with ideas ranging back as far as 1970s that are still influential in today's field. Bayesian optimization~\cite{movckus1974bayesian} was originally proposed to find global maxima of black box functions. This technique was successfully translated to estimating hyperparameters like learning rate, kernel width (SVMs), layer width, etc. by treating these as black box functions. Early years of classical deep learning saw the proposal of Grid search like algorithms~\cite{bergstra2012random, bergstra2011algorithms} to quickly estimate the best hyperparameters before training machine learning algorithms. However, as we show in this paper, blind application of effective techniques from machine learning is often computationally intensive owing to the high cost of function evaluation in QPUs.

\textbf{Quantum Fisher Information Matrix}: Quantum Fisher information matrix is a valuable tool in quantum sensing and metrology. Recently, QFIM has been used in the context of training variational algorithms by the QNG Optimizer~\cite{stokes2019quantum}. QNG optimizer offers better trainability and convergence as compared to ``classical optimizer". Furthermore, it has been established that computational resources required to compute QFIM are asymptotically insignificant when compared to resource requirements for training a VQA~\cite{van2021measurement}. The QFIM has been used to study expressiveness of quantum neural network models~\cite{abbas2020power}. Furthermore, QFIM has been used in the literature to suggest expressive and better trainable quantum circuits~\cite{haug2021capacity} and even find better learning rates for training VQAs~\cite{haug2021optimal}. Our proposed algorithm leveraging QFIM is thus another contribution in the rich body of work that exists in this domain and comes with the advantage of being rigorously motivated in theory. 

\textbf{Initializing distributions and barren plateaus}: Barren plateaus continue to be a significant challenge to overcome in the NISQ era VQAs. Several different remedies have been suggested in the literature to overcome these issues. One particular class of methods focuses on the role of initializing distributions. Different from the typical uniform initialization, the authors in~\cite{kulshrestha2022beinit} propose to use a beta distribution to delay the onset of barren plateaus in quantum circuits. Their results are empirically validated for QML style circuits in~\cite{herbst2024optimizing}. Zhang~\emph{et al}~\cite{zhang2022escaping} propose initializing with a Gaussian distribution where the variance scales as the inverse square root of the depth. Another sub-category of work aims to constrain both the ansatz and the initialization. Grant~\emph{et al}~\cite{grant2019initialization} propose to initialize parameters such that the unitary block evaluates to identity. A different approach is suggested in~\cite{park2023hamiltonian}. The authors in this work note that gradients do not exponentially vanish if they are constrained below a certain threshold $\tau_0$. They propose initializing parameters from a uniform distribution and constraining them below a pre-defined threshold. The various approaches to initializing distribution points to an underlying mechanism of action by which the ansatz is influenced. We hope that future works will attempt to derive such a framework that fits all existing results from these prior works.

\section{Conclusion}
In this paper, we motivated that one often overlooked inductive bias lies in the hyperparameters of the parameter initializing distribution in a VQA setting. To address the problem of finding such hyperparameters while keeping computational overhead low and remaining relevant to the ansatz and quantum task at hand, we proposed an algorithm for hyperparameter estimation. Our proposed algorithm is embarrassingly parallel and enjoys the advantage of being computationally efficient with increasing number of CPU cores. 

Through our numerical results, we empirically verify that our proposed algorithm leads to a consistently improved performance over manual hyperparameter selection for different parameterized initializing distributions and quantum tasks. We extend the numerical results to show that the hyperparameter selection can lead to a small improvement in gradient scaling for a two-design ansatz in some cases, but largely leaves the gradient scaling unchanged. This key result establishes the effectiveness of our algorithm since it is able to improve performance of an ansatz without leading it into a barren plateau. 

We hope that our paper inspires deeper investigation into the role of hyperparameters for different parametric distributions. We also hope that a deeper study be undertaken for studying QFIM and utilizing the information to design a better training protocol that is robust against barren plateaus.

% In this work we empirically motivate the importance of selecting appropriate hyperparameters for a given initializing distribution. We then provide a performant algorithm that finds optimial hyperparameter given a specific PQC. Our numerical results confirm that our hyperparameter selection algorithm leads to better initial parameters that enhance performance and rate of convergence in PQC.  

% This work was motivated by the empirical insight of the effect of hyperparameters on the gradient landscape produced by a given PQC. 

\bibliographystyle{IEEEtran}
\bibliography{ref, quant_refs}

\begin{thebibliography}{10}
\providecommand{\url}[1]{#1}
\csname url@samestyle\endcsname
\providecommand{\newblock}{\relax}
\providecommand{\bibinfo}[2]{#2}
\providecommand{\BIBentrySTDinterwordspacing}{\spaceskip=0pt\relax}
\providecommand{\BIBentryALTinterwordstretchfactor}{4}
\providecommand{\BIBentryALTinterwordspacing}{\spaceskip=\fontdimen2\font plus
\BIBentryALTinterwordstretchfactor\fontdimen3\font minus \fontdimen4\font\relax}
\providecommand{\BIBforeignlanguage}[2]{{%
\expandafter\ifx\csname l@#1\endcsname\relax
\typeout{** WARNING: IEEEtran.bst: No hyphenation pattern has been}%
\typeout{** loaded for the language `#1'. Using the pattern for}%
\typeout{** the default language instead.}%
\else
\language=\csname l@#1\endcsname
\fi
#2}}
\providecommand{\BIBdecl}{\relax}
\BIBdecl

\bibitem{mcclean2018barren}
\BIBentryALTinterwordspacing
J.~R. McClean, S.~Boixo, V.~N. Smelyanskiy, R.~Babbush, and H.~Neven, ``Barren plateaus in quantum neural network training landscapes,'' \emph{Nature communications}, vol.~9, no.~1, p. 4812, 2018. [Online]. Available: \url{https://www.nature.com/articles/s41467-018-07090-4}
\BIBentrySTDinterwordspacing

\bibitem{grant2019initialization}
\BIBentryALTinterwordspacing
E.~Grant, L.~Wossnig, M.~Ostaszewski, and M.~Benedetti, ``An initialization strategy for addressing barren plateaus in parametrized quantum circuits,'' \emph{Quantum}, vol.~3, p. 214, 2019. [Online]. Available: \url{https://quantum-journal.org/papers/q-2019-12-09-214/}
\BIBentrySTDinterwordspacing

\bibitem{kulshrestha2022beinit}
A.~Kulshrestha and I.~Safro, ``{BEINIT}: Avoiding barren plateaus in variational quantum algorithms,'' in \emph{2022 IEEE International Conference on Quantum Computing and Engineering (QCE)}.\hskip 1em plus 0.5em minus 0.4em\relax IEEE, 2022, pp. 197--203.

\bibitem{zhang2022escaping}
K.~Zhang, L.~Liu, M.-H. Hsieh, and D.~Tao, ``Escaping from the barren plateau via gaussian initializations in deep variational quantum circuits,'' \emph{Advances in Neural Information Processing Systems}, vol.~35, pp. 18\,612--18\,627, 2022.

\bibitem{park2023hamiltonian}
\BIBentryALTinterwordspacing
C.-Y. Park and N.~Killoran, ``Hamiltonian variational ansatz without barren plateaus,'' \emph{Quantum}, vol.~8, p. 1239, 2024. [Online]. Available: \url{https://quantum-journal.org/papers/q-2024-02-01-1239/}
\BIBentrySTDinterwordspacing

\bibitem{kandala2017hardware}
\BIBentryALTinterwordspacing
A.~Kandala, A.~Mezzacapo, K.~Temme, M.~Takita, M.~Brink, J.~M. Chow, and J.~M. Gambetta, ``Hardware-efficient variational quantum eigensolver for small molecules and quantum magnets,'' \emph{Nature}, vol. 549, no. 7671, p. 242–246, 2017. [Online]. Available: \url{https://www.nature.com/articles/nature23879}
\BIBentrySTDinterwordspacing

\bibitem{petz2011introduction}
D.~Petz and C.~Ghinea, ``Introduction to quantum fisher information,'' in \emph{Quantum probability and related topics}.\hskip 1em plus 0.5em minus 0.4em\relax World Scientific, 2011, pp. 261--281.

\bibitem{bergstra2011algorithms}
J.~Bergstra, R.~Bardenet, Y.~Bengio, and B.~K{\'e}gl, ``Algorithms for hyper-parameter optimization,'' \emph{Advances in neural information processing systems}, vol.~24, 2011.

\bibitem{salimans2017evolution}
T.~Salimans, J.~Ho, X.~Chen, S.~Sidor, and I.~Sutskever, ``Evolution strategies as a scalable alternative to reinforcement learning,'' \emph{arXiv preprint arXiv:1703.03864}, 2017.

\bibitem{wierstra2014natural}
D.~Wierstra, T.~Schaul, T.~Glasmachers, Y.~Sun, J.~Peters, and J.~Schmidhuber, ``Natural evolution strategies,'' \emph{The Journal of Machine Learning Research}, vol.~15, no.~1, pp. 949--980, 2014.

\bibitem{sun2009efficient}
Y.~Sun, D.~Wierstra, T.~Schaul, and J.~Schmidhuber, ``Efficient natural evolution strategies,'' in \emph{Proceedings of the 11th Annual conference on Genetic and evolutionary computation}, 2009, pp. 539--546.

\bibitem{bergholm2018pennylane}
\BIBentryALTinterwordspacing
V.~Bergholm, J.~Izaac, M.~Schuld, C.~Gogolin, M.~S. Alam, S.~Ahmed, J.~M. Arrazola, C.~Blank, A.~Delgado, S.~Jahangiri \emph{et~al.}, ``Pennylane: Automatic differentiation of hybrid quantum-classical computations,'' \emph{arXiv preprint arXiv:1811.04968}, 2018. [Online]. Available: \url{https://arxiv.org/abs/1811.04968}
\BIBentrySTDinterwordspacing

\bibitem{kingma2014adam}
D.~P. Kingma and J.~Ba, ``Adam: A method for stochastic optimization,'' \emph{arXiv preprint arXiv:1412.6980}, 2014.

\bibitem{movckus1974bayesian}
J.~Mo{\v{c}}kus, ``On bayesian methods for seeking the extremum,'' in \emph{IFIP Technical Conference on Optimization Techniques}.\hskip 1em plus 0.5em minus 0.4em\relax Springer, 1974, pp. 400--404.

\bibitem{bergstra2012random}
J.~Bergstra and Y.~Bengio, ``Random search for hyper-parameter optimization.'' \emph{Journal of machine learning research}, vol.~13, no.~2, 2012.

\bibitem{stokes2019quantum}
\BIBentryALTinterwordspacing
J.~Stokes, J.~Izaac, N.~Killoran, and G.~Carleo, ``Quantum natural gradient,'' \emph{Quantum}, vol.~4, p. 269, 2020. [Online]. Available: \url{https://quantum-journal.org/papers/q-2020-05-25-269/}
\BIBentrySTDinterwordspacing

\bibitem{van2021measurement}
B.~van Straaten and B.~Koczor, ``Measurement cost of metric-aware variational quantum algorithms,'' \emph{PRX Quantum}, vol.~2, no.~3, p. 030324, 2021.

\bibitem{abbas2020power}
\BIBentryALTinterwordspacing
A.~Abbas, D.~Sutter, C.~Zoufal, A.~Lucchi, A.~Figalli, and S.~Woerner, ``The power of quantum neural networks,'' \emph{arXiv preprint arXiv:2011.00027}, 2020. [Online]. Available: \url{https://arxiv.org/abs/2011.00027}
\BIBentrySTDinterwordspacing

\bibitem{haug2021capacity}
T.~Haug, K.~Bharti, and M.~Kim, ``Capacity and quantum geometry of parametrized quantum circuits,'' \emph{PRX Quantum}, vol.~2, no.~4, p. 040309, 2021.

\bibitem{haug2021optimal}
T.~Haug and M.~Kim, ``Optimal training of variational quantum algorithms without barren plateaus,'' \emph{arXiv preprint arXiv:2104.14543}, 2021.

\bibitem{herbst2024optimizing}
S.~Herbst, V.~De~Maio, and I.~Brandic, ``On optimizing hyperparameters for quantum neural networks,'' in \emph{2024 IEEE International Conference on Quantum Computing and Engineering (QCE)}, vol.~1.\hskip 1em plus 0.5em minus 0.4em\relax IEEE, 2024, pp. 1478--1489.

\end{thebibliography}
\end{document}